\documentstyle[preprint,aps,prl]{revtex}

\begin{document}

\preprint{{\bf ETH-TH/95-15}}

\title{Low-Field Phase Diagram of Layered Superconductors: \break The
  Role of Electromagnetic Coupling}

\author{Gianni Blatter$^{a}$, Vadim Geshkenbein$^{a,\, b}$, Anatoli
  Larkin$^{a,\, b}$, and Henrik Nordborg$^{a}$}

\address{$^{a\,}$Theoretische Physik, ETH-H\"onggerberg, CH-8093
  Z\"urich, Switzerland}

\address{$^{b\,}$L. D. Landau Institute for Theoretical Physics,
  117940 Moscow, Russia}

\date{\today}
\maketitle

\vspace*{-1.0truecm}
\begin{abstract}

  We determine the position and shape of the melting line in a layered
  superconductor taking the electromagnetic coupling between layers
  into account. In the limit of vanishing Josephson coupling we obtain
  a new generic reentrant low-field melting line.  Finite Josephson
  coupling pushes the melting line to higher temperatures and fields
  and a new line shape $B_{{\rm m}} \propto (1-T/T_c)^{3/2}$ is found.
  We construct the low-field phase diagram including melting and
  decoupling lines and discuss various experiments in the light of our
  new results.

\end{abstract}

Since its proposal in 1988\cite{Nelson}, vortex-lattice melting in
bulk type II material has become a central topic in the phenomenology
of high temperature superconductors. The order, position, and shape of
the transition have been investigated theoretically\cite{theory} as
well as experimentally\cite{experiment} by a large number of authors.
Most recently, the main interest is concentrating on the phase diagram
of the strongly layered Bi$_2$Sr$_2$Ca$_1$Cu$_2$O$_8$ (BiSCCO)
superconductor which is being investigated by means of
$\mu$SR\cite{muSR}, neutron scattering\cite{neutrons}, SQUID
magnetometry\cite{Pastoriza}, and Hall-sensor arrays\cite{Zeldov},
probing the melting- and/or decoupling transition in these materials.
It turns out that the most interesting regime is the low-field part of
the phase diagram with $B < 1$ kG, where the electromagnetic
interactions between the layers becomes relevant, and it is the
purpose of this letter to derive and analyze the vortex-lattice
melting transition in this regime, taking full account of
electromagnetic coupling.

The importance of electromagnetic interactions, contributing to the
stiffness of individual vortex lines, has been realized before within
the context of vortex-lattice melting in the dilute limit\cite{FFH},
where the transition line exhibits a reentrant behavior (lower branch
of the melting line). As we will show below, the electromagnetic
interaction also influences the behavior of the upper branch of the
low-field melting line and even may change its shape from the usual
$B_{{\rm m}}^{{\rm J}}(T) \propto (1-T/T_c)^2$ behavior to a new
power-law $B_{{\rm m}}^{{\rm em,J}}(T) \propto (1-T/T_c)^{3/2}$ within
a large part of the phase diagram --- this is the new and central
result of this paper.

Our analysis below is based on the continuum elastic description of
the vortex lattice combined with the Lindemann criterion, stating that
the lattice will undergo a melting transition once the mean thermal
displacement $\langle u^2 \rangle_{{\rm th}}^{1/2}$ becomes comparable
to the lattice spacing $a_\circ \approx (\Phi_\circ / B)^{1/2}$, $
\langle u^2 \rangle_{{\rm th}}^{1/2}/a_\circ \mid_{T_m,B_m} \approx
c_{{\rm \scriptscriptstyle L}}$.  The Linde\-mann number $c_{{\rm
\scriptscriptstyle L}}$ is usually chosen to be a constant of
order $c_{{\rm \scriptscriptstyle L}} \approx 0.1 - 0.3$.  Though not
rigorous, the Lindemann-type melting scenario has proven very useful
and reasonably accurate in predicting the positions of first-order
melting transitions in general and the line shape of the
vortex-lattice melting transition in particular.

A well known limiting case, where strong fluctuations due to
dimensional reduction drive a vortex-lattice melting transition, is
the superconducting film (2D dislocation-mediated Kosterlitz-Thouless
melting scenario, see \cite{HN}) and we will begin our analysis with
this elementary building block of a layered superconductor. Next, we
consider a layered system with electromagnetic coupling and derive the
shape of the reentrant melting line in this limit. Finally, we account
for the Josephson interaction between the layers producing a finite
anisotropy parameter $\varepsilon^2 = m/M <1$, where $m$ and $M$
denote the effective in-plane and $c$-axis masses.  Our results are
illustrated in Fig. 1, where we show the shape of the vortex-lattice
melting line as it evolves from the 2D isolated layer, to the
electromagnetically coupled system of layers, to the Josephson coupled
bulk anisotropic superconductor.

Our main task is the calculation of the mean-squared thermal
displacement\cite{review}
\begin{eqnarray}
\langle u^2 \rangle_{{\rm th}} \approx \int \frac{d^3 k}{(2\pi)^3}
\frac{T}{c_{66}K^2 + c_{44}({\bf k})k_z^2},
\label{usq}
\end{eqnarray}
with the shear modulus $c_{66}$ given by
\begin{equation}
c_{66} = \left\{ \begin{array}{r@{\quad\quad}l}
\displaystyle{\sqrt{\frac{\pi}{6}\frac{\lambda}{a_\circ}}
\frac{\varepsilon_\circ}{\lambda^2} e^{-a_\circ/\lambda}},
& \lambda < a_\circ,\\ \noalign{\vskip 5 pt}
\displaystyle{\frac{\varepsilon_\circ}{4 a_\circ^2},
\qquad\qquad\quad}
& a_\circ < \lambda,\end{array}\right.
\label{shear}
\end{equation}
and the dispersive tilt modulus $c_{44}({\bf k})$ consisting of a bulk
term $c_{44}^\circ ({\bf k})$ and a single vortex contribution
$c_{44}^c(k_z)$, $c_{44}({\bf k}) = c_{44}^\circ({\bf k}) +
c_{44}^c(k_z)$, with\cite{GK}
\begin{eqnarray}
c_{44}^\circ ({\bf k}) &=& \frac{\varepsilon_\circ}{a_\circ^2}
\frac{4 \pi \lambda^2/a_\circ^2} {1+(\lambda^2/\varepsilon^2) K^2
+ \lambda^2 k_z^2},
\label{tiltb} \\
c_{44}^c (k_z) &\approx& \frac{\varepsilon_\circ}{2 a_\circ^2}
\biggl[\varepsilon^2 {{\rm ln}} \biggl(\frac{\lambda^2/\varepsilon^2
\xi^2}{1+(\lambda^2/\varepsilon^2) K_{{\rm \scriptscriptstyle BZ}}^2
+ \lambda^2 k_z^2}\biggr)\nonumber\\
& &\quad\quad +\frac{1}{\lambda^2 k_z^2}{{\rm ln}}
\biggl(1+\frac{\lambda^2 k_z^2} {1+\lambda^2
K_{{\rm \scriptscriptstyle BZ}}^2}\biggr)\biggr]
\label{tiltsv}
\end{eqnarray}
(in (\ref{usq}) we neglect a second contribution to $\langle u^2
\rangle_{{\rm th}}$ involving lattice compression and keep only the
main term).  Here, $\varepsilon_\circ = (\Phi_\circ / 4\pi\lambda)^2$
denotes the basic energy scale of the continuum elastic theory,
$\Phi_\circ = hc/2e$ is the flux quantum, $\lambda$ denotes the planar
London penetration depth, and $\xi$ is the planar coherence length.
The second term in the single vortex tilt $c_{44}^c$ is due to the
electromagnetic coupling between the layers and is the only term in
$c_{44}$ surviving the limit $\varepsilon \rightarrow 0$ (layer
decoupling).  The electromagnetic contribution to the tilt modulus is
strongly dispersive and produces the large stiffness $\varepsilon_l
\approx \varepsilon_\circ/2$ of the vortex lines in the
long-wave-length limit $k_z < 1/\lambda$. With increasing $k_z$ the
electromagnetic stiffness decays $\propto 1/\lambda^2k_z^2 $ and the
line tension crosses over to the well known result $\varepsilon_l
\approx \varepsilon^2 \varepsilon_\circ$ for the anisotropic
superconductor as $k_z$ increases beyond $1/\varepsilon\lambda$ (note
that this residual tension is due to the Josephson coupling and is
relevant only for $\varepsilon \lambda > d$, where $d$ denotes the
layer separation).  The expression given in (\ref{tiltsv}) is valid
for small displacements, in the elastic regime. For large
displacements $u k_z > 1$ the logarithm in the second term of
(\ref{tiltsv}) should be cut on $2\lambda/u$ rather than $\lambda
k_z$\cite{clem}. In our analysis below we then replace the logarithm
by the factor $\lambda^2 k_z^2 / (1 + \beta \lambda^2 k_z^2)$ with
$\beta = 1/{{\rm ln}}(1+4\lambda^2/ c_{{\rm \scriptscriptstyle L}}^2
a_\circ^2)$ producing a smooth interpolation between the hard and soft
tilt modes at large and small wave-lengths, respectively.

We start with the analysis of an {\it individual layer} (we use the
definition $\lambda^2/d = \lambda_s^2/d_s$ with $\lambda_s$ and $d_s$
the penetration depth and thickness of the superconducting layer).
Dropping the tilt energy in (\ref{usq}), the integral over $k_z$
provides a factor $2\pi/d$ and cutting the $K$-integration on a few
lattice spacings we obtain the ratio $\langle u^2 \rangle_{{\rm th}}/
a_\circ^2 \approx T/2 \pi c_{66} d a_\circ^2$. Within our simple
Lindemann approach we then can reproduce the correct result $T_{{\rm
m}}^{{\rm \scriptscriptstyle 2D}} \approx a_\circ^2 d c_{66}/4\pi$
for the dislocation-mediated melting temperature if we choose a
Lindemann number $c_{{\rm\scriptscriptstyle L}} = 1/2\sqrt{2}\pi
\approx 0.1$. The high-field part ($a_\circ < \lambda$) of the melting
line is field-independent,
\begin{eqnarray}
T_{{\rm m}}^{{\rm \scriptscriptstyle 2D}} \approx
\frac{\varepsilon_\circ d}{16 \pi},
\label{ttd}
\end{eqnarray}
and using parameters typical for the layered high-$T_c$
superconductors, $T_c \approx 100$ K, $\lambda^2(T) \approx
\lambda_0^2/(1-T^2/T_c^2)$ with $\lambda_0 \approx 1800~\AA$, and $d =
15~\AA$, we obtain $\varepsilon_{\circ}(T=0) d \approx 10^3$ K and
$T_{{\rm m}}^{{\rm \scriptscriptstyle 2D}} \approx 20$ K.  The
low-field part ($a_\circ > \lambda$) of the melting line is dominated
by the exponential decay of the shear modulus and we obtain the result
\begin{eqnarray}
B_{{\rm m}}^{{\rm \scriptscriptstyle 2D}} \approx
\frac{\Phi_\circ}{\lambda^2} \biggl[{{\rm ln}} \biggl(
(2\pi^3/3)^{1/2} c_{{\rm\scriptscriptstyle L}}^2
\frac{\varepsilon_\circ d}{T} \biggr) \biggr]^{-2}.
\label{btd}
\end{eqnarray}
The result for the melting line of an isolated layer is illustrated in
Fig.\ 1.  Note that the result (\ref{btd}) is a somewhat artificial
construct, as we have used the low-field expression in
(\ref{shear}) for our analysis. In this way we illustrate the behavior
of the melting line in the absence of any interlayer coupling (neither
electromagnetic nor Josephson) while keeping the shear modulus of a
translation invariant system along the field axis. On the other hand,
the analysis of a 2D film involves a different shear modulus\cite{CS}
which softens only at very low fields ($a_\circ > \lambda_{{\rm eff}}
= 2 \lambda^2/d$) following the power-law behavior $c_{66} \approx
0.46 \, \varepsilon_\circ \lambda_{{\rm eff}}/a_\circ^3 \propto
B^{3/2}$ rather than the exponential behavior used in our analysis.

Next we consider a finite {\it electromagnetic coupling} between the
layers while keeping $\varepsilon = 0$ (no Josephson coupling). In the
high-field regime ($a_\circ < \lambda$) the shear term in (\ref{usq})
dominates over the tilt energy and we recover the field independent
2D-result (\ref{ttd}). For small fields with $a_\circ > \lambda$ the
tilt energy becomes relevant and the Lindemann criterion reads
\begin{eqnarray}
c_{{\rm\scriptscriptstyle L}}^2 \approx \frac{2 T}
{\varepsilon_\circ d}\frac{\lambda^2}{a_\circ^2}
\Biggr[\frac{1}{4\pi\delta} {{\rm ln}}(1 + 4 \pi \delta \beta) +
\frac{d}{\lambda (4 \pi\delta)^{1/2}}\Biggr],
\label{Lem}
\end{eqnarray}
with $\delta = 2 c_{66} \lambda^2 / \varepsilon_\circ$. Here, the
first term originates from the soft tilt modes with $k_z \lambda > 1$,
whereas the second term involves the long wave-length modes hardened
by the electromagnetic coupling. This second term becomes relevant
only at very small fields $a_\circ / \lambda \gg 1 $, where the shear
modulus is exponentially small, $\delta \propto
\exp(-a_\circ/\lambda)$.  The result (\ref{Lem}) provides a lower
branch of the melting line which is limited by soft shear and hard
tilt,
\begin{eqnarray}
B_{{\rm m}}^{{\rm em,l}} (T) \approx \frac{\Phi_\circ}{\lambda^2}
\frac{1}{4}\biggl[{{\rm ln}} \biggl(
\frac{4\pi c_{{\rm\scriptscriptstyle L}}^2}{(3\pi)^{1/4}}
\frac{\varepsilon_\circ \lambda}{T} \biggr)\biggr]^{-2},
\label{beml}
\end{eqnarray}
as well as a tilt limited upper branch
\begin{eqnarray}
B_{{\rm m}}^{{\rm em,u}} (T) \approx \frac{\Phi_\circ}{\lambda^2}
\frac{c_{{\rm\scriptscriptstyle L}}^2}{2\beta} \frac{
\varepsilon_\circ d}{T} \propto \biggl(1-\frac{T^2}{T_c^2}\biggr)^2.
\label{bemu}
\end{eqnarray}
The two branches merge near $T_c$,
\begin{eqnarray}
1 - \frac{T_x}{T_c} \approx \frac{\beta G^{{\rm
\scriptscriptstyle 2D}}}{4 c_{{\rm\scriptscriptstyle L}}^2}
\biggl[{{\rm ln}} \biggl(\frac{2\pi (2\beta)^{1/2}}{(3\pi)^{1/4}}
\frac{c_{{\rm\scriptscriptstyle L}}}{\sqrt{G^{{\rm
\scriptscriptstyle 2D}}}} \frac{\lambda_0}{d} \biggr) \biggr]^{-2},
\label{txem}
\end{eqnarray}
and no solid phase can exist at high temperatures beyond $T_x$. Using
typical parameters for the layered high-$T_c$ materials and adopting a
value $c_{{\rm\scriptscriptstyle L}} \approx 0.1$ for the Lindemann
number we find $T_x$ close to $T_c$, $1 - T_x/T_c \approx 0.05$ (in
(\ref{txem}) we have introduced the 2D Ginzburg number $G^{{\rm
\scriptscriptstyle 2D}} \approx T_c/\varepsilon_\circ(T=0) d
\approx 0.1$; the logarithms in (\ref{beml}) and (\ref{txem}) take
typical values around 5 -- 6). The reentrant melting line defined by
(\ref{beml}) and (\ref{bemu}) is illustrated in Fig. 1: The
electromagnetic coupling of the layers favors the solid phase and the
low-field melting line develops the characteristic ``nose-like'' shape
of a 3D system. Note that the point of reentrance ends up in the
critical region close to $T_c$. Since our approach accounts for the
fluctuations of the phase-field via the thermal motion of vortices but
neglects amplitude fluctuations of the order parameter our analysis
breaks down in this regime.

In the final step we account for the {\it Josephson coupling} between
the layers producing a finite anisotropy parameter $\varepsilon > 0 $.
This additional coupling becomes relevant whenever $a_\circ,\lambda >
d/\varepsilon$ and again favors the solid phase, hence pushing the
melting line further towards higher temperatures and fields.
Evaluating the Lindemann criterion in the low-field regime ($a_\circ >
\lambda$) we recover the previous result (\ref{Lem}) with the
modification that the soft tilt modes are cut off on $1/\varepsilon
\sqrt{\beta}\lambda$ instead of $\pi/d$, leading to the replacement of
${{\rm ln}(\dots)}/4 \pi \delta$ by $(d \sqrt{\beta}/\pi \varepsilon
\lambda) [{{\rm ln}}(\dots)/4 \pi \beta \delta + 1]$ in the first
term of (\ref{Lem}). The lower branch of the melting line remains
unaffected, whereas the upper branch of the low-field melting line
takes the new form
\begin{eqnarray}
B_{{\rm m}}^{{\rm em,J}} (T) \approx \frac{\Phi_\circ}{\lambda^2}
\frac{\pi c_{{\rm\scriptscriptstyle L}}^2}{4\sqrt{\beta}}
\frac{\varepsilon \varepsilon_\circ \lambda}{T} \propto
\biggl(1-\frac{T^2}{T_c^2}\biggr)^{3/2}.
\label{bemJ}
\end{eqnarray}
The crossing point of the lower and upper branch of the melting line
is shifted towards higher temperatures,
\begin{eqnarray}
1 - \frac{T_x}{T_c} \approx \frac{1}{2}\Biggl[\frac{\sqrt{\beta}
G^{{\rm \scriptscriptstyle 2D}}}{\pi c_{{\rm\scriptscriptstyle L}}^2}
\frac{d}{\varepsilon \lambda_\circ} \biggl[{{\rm ln}} \biggl(\frac{4
\sqrt{\beta}}{(3\pi)^{1/4}\varepsilon} \biggl) \biggr]^{-2}\Biggr]^2.
\label{txJ}
\end{eqnarray}
For $\varepsilon \lambda_0 < d$ the line $B_{{\rm m}}^{{\rm em,J}}$
goes over into the generic melting line $B_{{\rm m}}^{{\rm em,u}}$ as
the temperature drops below $T^{{\rm em}} \approx T_c [1 -
\beta(\varepsilon \pi \lambda_0/d)^2]^{1/2}$.  For the opposite case
where $\varepsilon \lambda_0 > d$ the generic line $B_{{\rm m}}^{{\rm
em,u}}$ is completely hidden and $B_{{\rm m}}^{{\rm em,J}}$ merges
into the well known bulk anisotropic melting line $B_{{\rm m}}^{{\rm
J}}$ as the field grows beyond $\Phi_\circ/\lambda^2$: At these
fields the tilt energies are dominated by the dispersive bulk term
$c_{44}^\circ \approx 4 \pi \varepsilon^2 \varepsilon_\circ/ a_\circ^4
K^2$ (see Eq. (\ref{tiltb})) and the Lindemann criterion provides the
well known result
\begin{eqnarray}
B_{{\rm m}}^{{\rm J}} (T) \approx \frac{\Phi_\circ}{\lambda^2}
4 \pi c_{{\rm\scriptscriptstyle L}}^4 \frac{\varepsilon^2
\varepsilon_\circ^2 \lambda^2}{T^2} \propto
\biggl(1-\frac{T^2}{T_c^2}\biggr)^2.
\label{bb}
\end{eqnarray}
At large fields ($a_\circ < d/\varepsilon < \lambda_\circ$) the 2D
result (\ref{ttd}) is recovered.

The most interesting result is the new line shape $B_{{\rm m}}^{{\rm
em,J}} \sim (1-T/T_c)^{3/2}$, Eq. (\ref{bemJ}), describing the
low-field/high-temperature melting in a Josephson-coupled layered or
highly anisotropic superconductor (small parameter $\varepsilon <
d/\lambda_\circ$). This new result is due to the electromagnetic
coupling which dominates over the bulk-dispersive tilt modulus
$c_{44}^\circ$ as well as over the single-vortex line tension
$\varepsilon^2 \varepsilon_\circ$ due to Josephson-coupling in this
regime. The substitution of the old result (\ref{bb}) by the new
expression (\ref{bemJ}) is particularly relevant in the strongly
layered superconductors such as BiSCCO: The $(1-T/T_c)^{3/2}$
power-law is valid provided that $d/\pi\sqrt{\beta}\varepsilon <
\lambda < a_\circ$.  Assuming $\varepsilon \sim 1/150$, the second
restriction implies $T > 0.4 \, T_c$.  In less anisotropic materials,
such as YBCO with $\varepsilon \approx 1/5$, this condition is much
more stringent and the upper branch of the melting line is always
described by the old result, Eq. (\ref{bb}). Note, however, that in
YBCO the suppression of the order parameter close to the upper
critical field $H_{c_2}$ becomes relevant and the melting line cannot
be described in terms of a simple power law $\propto (1-T/T_c)^2$ any
longer, see Ref. \cite{theory}, Blatter and Ivlev, for a detailed
discussion (in BiSCCO the melting line is far below $H_{c_2}$ and
there is no suppression of the order parameter in this regime).

It is instructive to compare the different low-field melting lines as
given by Eqs.\ (\ref{bemu}), (\ref{bemJ}), and (\ref{bb}). A quick
inspection gives the ratio $B_{{\rm m}}^{{\rm em,u}}/B_{{\rm m}}^{{\rm
J}} = (d/\lambda\varepsilon c_{{\rm \scriptscriptstyle L}})^2
T/8\pi\beta\varepsilon_\circ d$, which is of order unity taking the
above parameters for BiSCCO and using $\varepsilon = 1/150$, a value
often quoted in the literature\cite{muSR,neutrons}.  Similarly,
$B_{{\rm m}}^{{\rm em,J}}/B_{{\rm m}}^{{\rm J}} =
(d/\lambda\varepsilon c_{{\rm \scriptscriptstyle L}}) T/16
\sqrt{\beta} c_{{\rm \scriptscriptstyle L}}\varepsilon_\circ d \approx
\alpha [T^2/T_c(T_c-T)]^{1/2}$, were again $\alpha \sim 1$ if we use
the above parameters for BiSCCO. The comparison of experimental data
for the irreversibility or melting line with the theoretical
prediction is often used to extract an estimate of the anisotropy
parameter $\varepsilon$, particularly in the strongly layered
materials\cite{muSR,neutrons}.  Following up the above discussion we
draw attention to an important problem with this procedure: If the
anisotropy parameter is very small, say $\varepsilon < 1/500$, the
(upper branch of the) low-field melting line (where the comparison
theory/experiment is carried out) is dominated by the electromagnetic
coupling and no anisotropy parameter can be extracted. The analysis of
the melting line can provide a reliable estimate for the anisotropy
parameter only if $\varepsilon$ is large enough such that either the
bulk result (\ref{bb}) is valid or the mixed electromagnetic/Josephson
result (\ref{bemJ}) can be identified via its particular line shape.

It is generally believed that the low-field melting transition takes
the vortex lattice into a liquid of vortex {\it lines}. It then has
been proposed that this line-liquid transforms into a pancake-liquid
in a second transition where the layers decouple, see Refs.
\cite{GK,review}. A simple estimate for the position and shape of this
decoupling line is obtained in the following way: thermal wandering of
the vortex line over a distance $L$ produces a displacement amplitude
$\langle u^2\rangle_{{\rm th}} \simeq L T/\varepsilon_l$. The layers
decouple when the mean thermal displacement between line segments in
neighboring layers becomes of the order of the lattice spacing,
$\langle u^2\rangle_{{\rm th}}^{1/2} (L=d) \simeq a_\circ$.  For a
Josephson-coupled system the line tension is $\varepsilon_l \simeq
\varepsilon^2 \varepsilon_\circ$ and we obtain the well known result
$B_{{\rm dc}}^{{\rm\scriptscriptstyle J}} \simeq
[\Phi_\circ/(d/\varepsilon)^2] \, \varepsilon_\circ d/T \propto
1-T/T_c$.  However, for small anisotropy the electromagnetic coupling
dominates at low fields and using the short wave-length elasticity
$\varepsilon_l \simeq \varepsilon_\circ (d/\lambda)^2$ we find that
the decoupling line follows the melting line $B_{{\rm m}}^{{\rm
em,u}}$. We then obtain a phase diagram where the decoupling and
melting lines are separate transitions at low ($T < T_{{\rm m}}^{{\rm
\scriptscriptstyle 2D}}$) and high ($T > T^{{\rm em}}$)
temperatures but close up in between.

Recently, a first-order phase transition has been observed in the
low-field regime of a strongly layered BiSCCO
superconductor\cite{Zeldov}. The jump in the magnetization observed at
the transition can be associated either with a vortex-lattice melting-
or with a layer-decoupling transition. Fits using a $(1-T/T_c)^{1.55}$
(melting) or a $(T_c/T - 1)$ (decoupling) power-law behavior produce a
satisfactory agreement with the data over most of the measured
temperature interval\cite{Zeldov}. Our new result (\ref{bemJ}) then is
in good agreement with the measured power-law behavior based on the
melting scenario. Whether the observed transition indeed can be
attributed to a first-order melting transition remains to be shown,
however.

In conclusion, we have determined the position and shape of the
melting line in a layered superconductor taking the electromagnetic
interaction between the layers into account. Whereas the
electromagnetic coupling is irrelevant at fields $B >
\Phi_\circ/\lambda^2$, new results for the melting line have been
obtained in the low-field regime $B < \Phi_\circ / \lambda^2$. In this
regime, the electromagnetic coupling produces a stiffening of the
vortex line at long wave lengths $k_z < 1/\lambda$.  Both, the lower
and the upper branch of the reentrant melting line are effected by
this stiffening and a characteristic ``nose''-shaped 3D melting line
is found even in the absence of Josephson coupling between the layers.
Accounting for an additional Josephson coupling, the upper branch of
the melting line is pushed out to higher temperatures and fields and
takes on a new characteristic line shape $\propto (1-T/T_c)^{3/2}$, as
observed recently in a BiSCCO superconductor\cite{Zeldov}. The results
are crucial for an accurate understanding of the low-field phase
diagram of layered superconductors.

We thank E. H. Brandt for helpful discussions and the Swiss National
Foundation for financial support.

\begin{figure}
\caption{
  Low-field phase diagram of a strongly layered superconductor.
  Reduced units $b = B/(\Phi_\circ/\lambda_\circ^2)$ and $t = T/T_c$
  have been used and the Lindemann number $c_{{\rm \scriptscriptstyle
  L}} = 0.1$ has been chosen (else parameters appropriate for
  BiSCCO have been adopted, see text). The thick lines show the
  results for the isolated 2D layer and for the electromagnetically
  coupled system. The thin lines incorporate the effect of a finite
  Josephson coupling between the layers for anisotropy parameters
  $\varepsilon = 1/500$, $1/150$, and $1/50$. The inset shows the same
  results on a logarithmic field scale, where the reentrant behavior
  of the melting line becomes more visible. The dotted line traces
  $b(t) = B/[\Phi_\circ/\lambda^2(t)]$.}
\label{fig:1}
\end{figure}
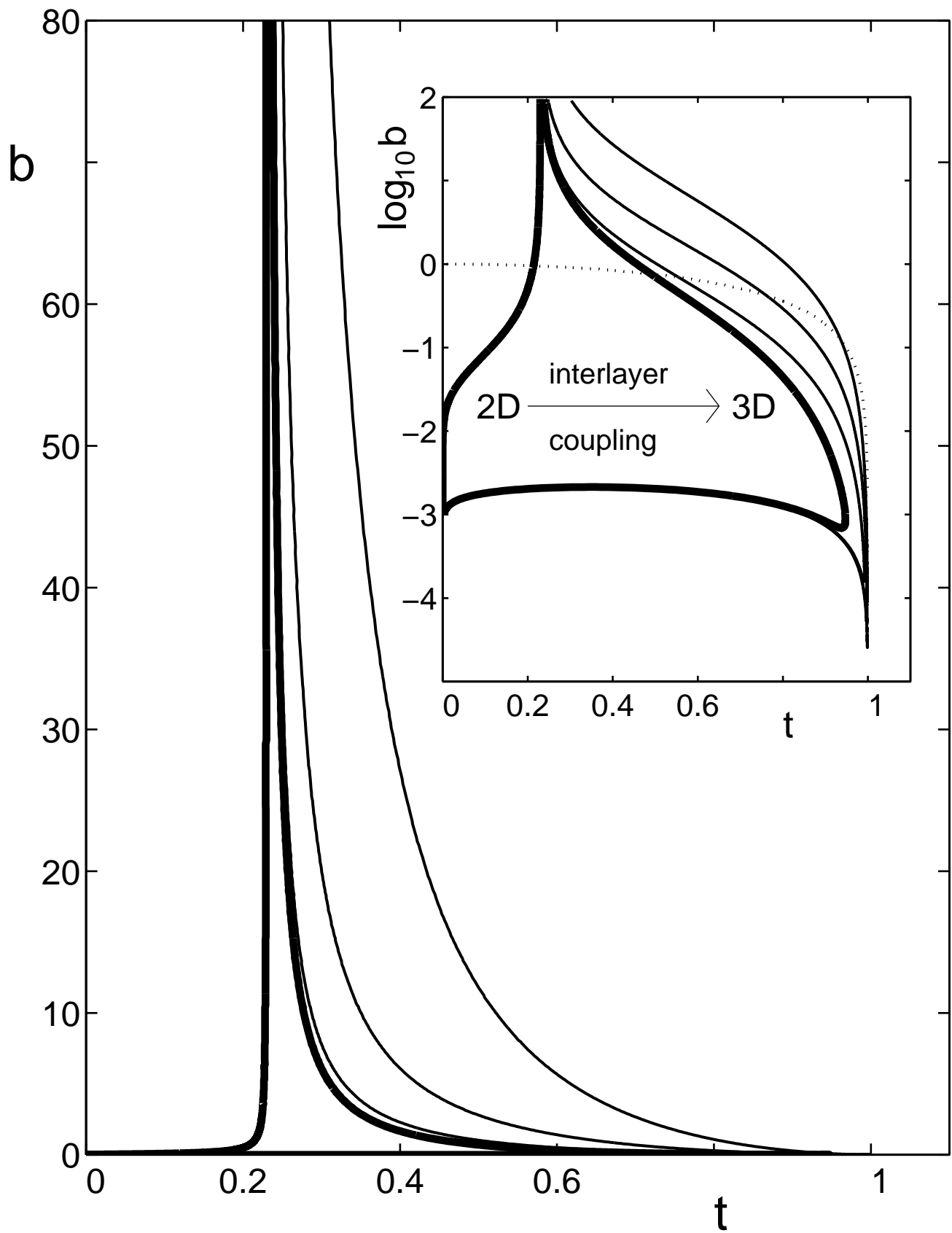
\end{document}